\documentclass{elsart}

\usepackage{epsfig}
\begin{document}
\begin{frontmatter}
%
%
%
\title{Correlations of errors in a CP-violation neutrino factory experiment}
%
%
\author{John Pinney}
\address{The Daiwa Anglo-Japanese Foundation\\Daiwa Securities Kabuto-cho Bldg 9F\\1-1-9 Nihonbashi Kayabacho, Chuo-ku, Tokyo 103-0025, Japan}
\begin{abstract}
An indirect CP-violation neutrino factory experiment is studied, including simultaneous correlations of errors of the six oscillation parameters and a matter effect normalisation parameter, $C$. It is found that for larger values of $\theta_{13}$ ($3^\circ$), an uncertainty on the matter effect of the order 5\% means that the optimal baseline for measurement of $\delta$ is reduced from 3000km to 2000km. When the background fraction is assumed to be $10^{-3}$ the optimal muon energy is also reduced from 50GeV to around 20GeV. An experiment with $L=2000$km, $E_\mu=$20GeV is found to be sensitive to non-zero $\delta$ down to around $\theta_{13}\approx0.3^\circ$.
\end{abstract}
\end{frontmatter}
\section{Introduction}
The ultimate aim of the next generation of neutrino experiments has to be the measurement of the CP phase of the MNS mixing matrix, $\delta$. Assuming that high-energy electrons and positrons cannot be distinguished by detectors, our only source of information in a neutrino factory experiment is the number of wrong- and right-sign muons, corresponding (for neutrinos from muon decay) to oscillation of $\bar{\nu_e}$ and no oscillation of $\nu_\mu$ respectively. It has recently been realised that uncertainties in the electron density along the neutrino flight path may be sufficient to cause problems at long baselines and high energies. This talk is a summary of our recent work towards resolving the issue of optimal baseline and muon energy for the measurement of $\delta$ at a neutrino factory \cite{corr}.
\section{Hypothesis testing}
We considered an indirect CP-violation experiment with $10^{21}$ useful muon decays and $10^{21}$ antimuon decays\footnote{Note that despite the difference in cross-section of neutrinos and antineutrinos, if an explicit demonstration of CP violation is not the aim of the experiment then it is not necessarily advantageous to run for twice as long with muons as antimuons in the storage ring. For a fixed time period, the most efficient proportions of muon to antimuon running are determined by maximising the overall number of wrong-sign muons and antimuons whilst retaining sufficient wrong-sign antimuons to give significant data in that channel. We found that equal numbers of muon and antimuon decays gave slightly better statistics for the oscillation parameters studied than the usual 2:1 ratio of muons to antimuons.}. Taking into account all of the channels $\nu_e\rightarrow \nu_\mu$, $\bar{\nu_e}\rightarrow\bar{\nu_\mu}$, $\nu_\mu\rightarrow \nu_\mu$ and $\bar{\nu_\mu}\rightarrow\bar{\nu_\mu}$, we examined whether or not a hypothesis of CP phase equal to $\bar{\delta}$ is rejected when the true value is $\delta$. To test this hypothesis, we used a parameter $\Delta\chi^2$ as defined in \cite{corr}. 
The electron number density as a function of distance from the source, $N_{e}(x)$, is an input for the oscillation probabilities but is not known exactly. The current best estimate of the density of the Earth as a function of radius is the Preliminary Reference Earth Model, derived from seismological data \cite{PREM}. However, it is known that there are sizeable latitude-longitude dependent fluctuations in matter density and composition in the crust and upper mantle \cite{geller}, so local deviations from this spherically-symmetric model may be considerable. To investigate the effect of the uncertainty in electron number density on the usefulness of an experiment, we introduced a parameter $C$, the overall normalisation of the PREM. Hence 
$\Delta\chi^2$ is a function of six oscillation parameters (3 times $\theta_{ij}$, 2 times $\Delta m^2_{ij}$ and $\delta$), as well as $C$, where
$C=1$ corresponds to the model as given and fluctuations away from the PREM can be represented by changing the value of $C$. The background rate $f_B$ and systematic error on the background $\sigma_B$ were also included in the analysis.
\vglue -1.5cm
\section{Minimisation of $\Delta\chi^2$}
To find the optimum energy and baseline, we used the quantity $\Delta\chi^2_{min}$, defined as the minimum of $\Delta\chi^2$ when we simultaneously vary the mixing angles and mass-squared differences within the region allowed at the 90\% confidence level by the atmospheric and solar neutrino data \cite{corr}. The matter effect normalisation $C$ was varied by 5\% between 0.95 and 1.05. This minimum was found numerically using a `downhill simplex' method \cite{recipes}. Using $\Delta\chi^2_{min}$, we found the minimum data size $D$, expressed in kt$\cdot$yr, which is needed to reject a hypothesis $\bar{\delta}=0$ at the 3$\sigma$ confidence level when the true value of $\delta$ is $\pi/2$.
\begin{figure}[p]
\begin{center}
\resizebox{13cm}{!}{
\rotatebox{270}{
\includegraphics{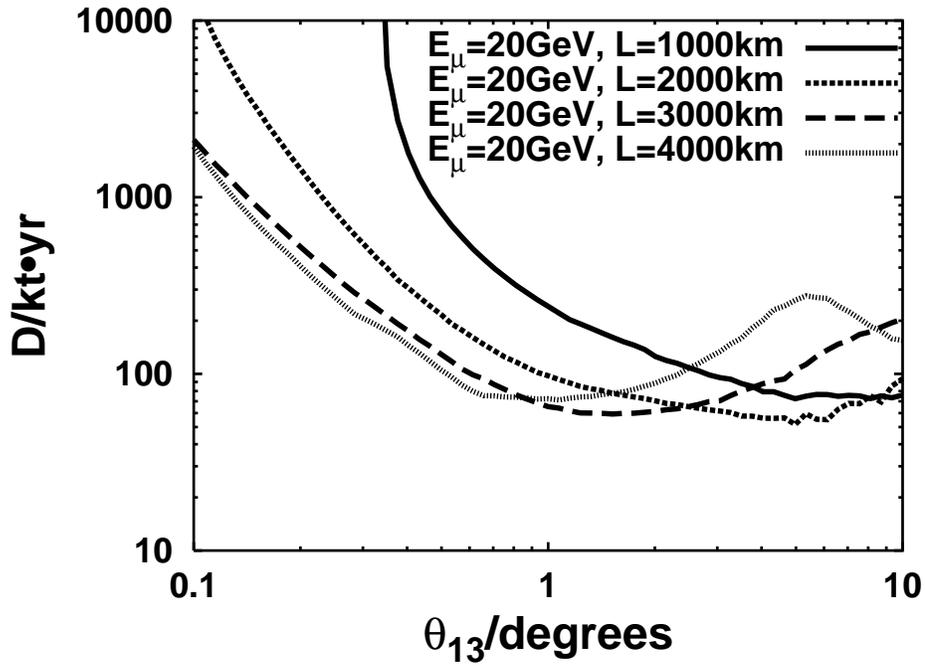}
}
}
\vglue 1.0cm
\resizebox{13cm}{!}{
\rotatebox{270}{
\includegraphics{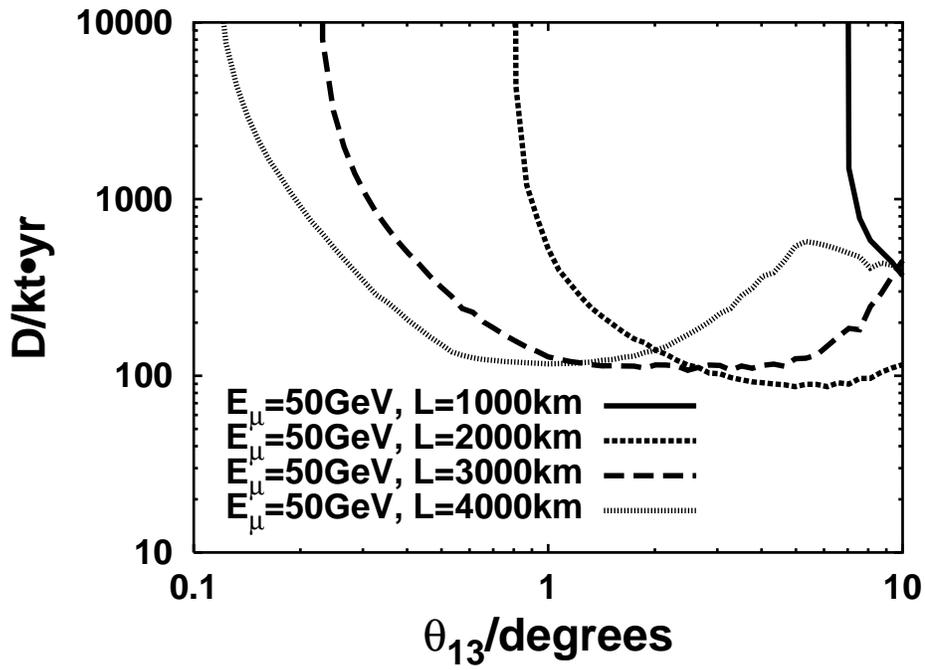}
}
}
\vglue 1.0cm
\caption{Data size (kt$\cdot$yr) required to reject a hypothesis of $\bar{\delta}=0$ at 3$\sigma$ when the true value is $\delta=\pi/2$, in the case of a neutrino factory with $10^{21}$ useful muon decays per year and a background fraction $f_B=10^{-3}$.}
\end{center}
\end{figure}
\section{Conclusions}
The necessary data size as a function of $L$ and $E_\mu$ was found to have a minimum of around 60 kt$\cdot$yr, but the optimal baseline and muon energy varied considerably with the values of $\delta$, $\theta_{13}$, $f_B$ and the threshold energy for neutrino detection, $E_{th}$. At very high energies, the contribution of the systematic error to the total error becomes large, and sensitivity is lost. Also, the correlation ($\delta\leftrightarrow C$) becomes strong at long baselines, so the optimal $L$ is shorter than might otherwise be expected. The set of parameters $L=$2000km, $E_\mu=$20GeV gave the largest range of sensitivity to $\delta$, down to $\theta_{13}\approx0.3^\circ$ for $f_B=10^{-3}$, $E_{th}=0.1$GeV and a 500kt detector, as shown in Fig. 1.
\end{document}